\documentclass[floatfix, noshowpacs, preprintnumbers, twocolumn, amsmath, amssymb, aps, superscriptaddress, prb]{revtex4}

\pdfoutput=1

\usepackage{graphicx, xcolor}
\usepackage[caption=false]{subfig}
\usepackage{soul}
\usepackage{mathrsfs}

\newcommand{\bra}[1]{\left< #1 \right|} 
\newcommand{\ket}[1]{\left| #1 \right>}

\begin{document}

\title{TensorFlow Solver for Quantum PageRank in Large-Scale Networks}

\author{Hao Tang}
\affiliation{Center for Integrated Quantum Information Technologies (IQIT), School of Physics and Astronomy and State Key Laboratory of Advanced Optical Communication Systems and Networks, Shanghai Jiao Tong University, Shanghai 200240, China}
\affiliation{CAS Center for Excellence and Synergetic Innovation Center in Quantum Information and Quantum Physics, University of Science and Technology of China, Hefei, Anhui 230026, China}

\author{Tian-Shen He}
\affiliation{Center for Integrated Quantum Information Technologies (IQIT), School of Physics and Astronomy and State Key Laboratory of Advanced Optical Communication Systems and Networks, Shanghai Jiao Tong University, Shanghai 200240, China}

\author{Ruo-Xi Shi}
\affiliation{Center for Integrated Quantum Information Technologies (IQIT), School of Physics and Astronomy and State Key Laboratory of Advanced Optical Communication Systems and Networks, Shanghai Jiao Tong University, Shanghai 200240, China}

\author{Yan-Yan Zhu}
\affiliation{School of Physical Science, University of Chinese Academy of Science, Beijing 100049, China}

\author{Marcus Lee}
\affiliation{Department of Physics, Cambridge University, Cambridge CB3 0HE, UK}

\author{Tian-Yu Wang}
\affiliation{Center for Integrated Quantum Information Technologies (IQIT), School of Physics and Astronomy and State Key Laboratory of Advanced Optical Communication Systems and Networks, Shanghai Jiao Tong University, Shanghai 200240, China}

\author{Xian-Min Jin}
\email{xianmin.jin@sjtu.edu.cn} 
\affiliation{Center for Integrated Quantum Information Technologies (IQIT), School of Physics and Astronomy and State Key Laboratory of Advanced Optical Communication Systems and Networks, Shanghai Jiao Tong University, Shanghai 200240, China}
\affiliation{CAS Center for Excellence and Synergetic Innovation Center in Quantum Information and Quantum Physics, University of Science and Technology of China, Hefei, Anhui 230026, China}
\email{xianmin.jin@sjtu.edu.cn} %% email address is required

\maketitle
\textbf{Google PageRank is a prevalent and useful algorithm for ranking the significance of nodes or websites in a network, and a recent quantum counterpart for PageRank algorithm has been raised to suggest a higher accuracy of ranking comparing to Google PageRank. The quantum PageRank algorithm is essentially based on quantum stochastic walks and can be expressed using Lindblad master equation, which, however, needs to solve the Kronecker products of an O($N^4$) dimension and requires severely large memory and time when the number of nodes $N$ in a network increases above 150. Here, we present an efficient solver for quantum PageRank by using the Runge-Kutta method to reduce the matrix dimension to O($N^2$) and employing TensorFlow to conduct GPU parallel computing. We demonstrate its performance in solving quantum PageRank for the USA major airline network with up to 922 nodes. Compared with the previous quantum PageRank solver, our solver dramatically reduces the required memory and time to only 1\% and 0.2\%, respectively, making it practical to work in a normal computer with a memory of 4-8 GB in no more than 100 seconds. This efficient solver for large-scale quantum PageRank and quantum stochastic walks would greatly facilitate studies of quantum information in real-life applications. }

%\section{Introduction}
Navigation through the World Wide Web (WWW) has nowadays become an indispensable way to obtain information in everyday life. A dramatically growing number of webpages that contain various information has revealed the urgent demand of an effective tool to sort and rank the webpages for effective information searching. The PageRank raised by Google is a most representative and successful example to accomplish such tasks\cite {Brin1998, Page1999}. The key concept for Google PageRank is to treat the entire internet as a directed graph, where each website can be regarded a node of the graph, and each hyperlink that directs website navigators from one website to another is treated as an edge. After long-time stochastic navigation in the website network which is essentially a classical random walk process, there would eventually be a stable probability distribution and by sorting the websites according to their probability values, one would get the ranking of significance for these websites. PageRank has also been wide applied to a larger diversity of networks. For instance, evaluating the impact of a scientist through his connections in the academic network\cite{Radicchi2009}, studying species within an ecosystem \cite{Allesina2009} , and finding key neurons in the neural network of the worm C. $elegans$ \cite{Watts1998,Sanchezburillo2012} can all be reduced to the model of element ranking that PageRank manages to do. 

\begin{figure*}[ht!]
\includegraphics[width=0.92\textwidth]{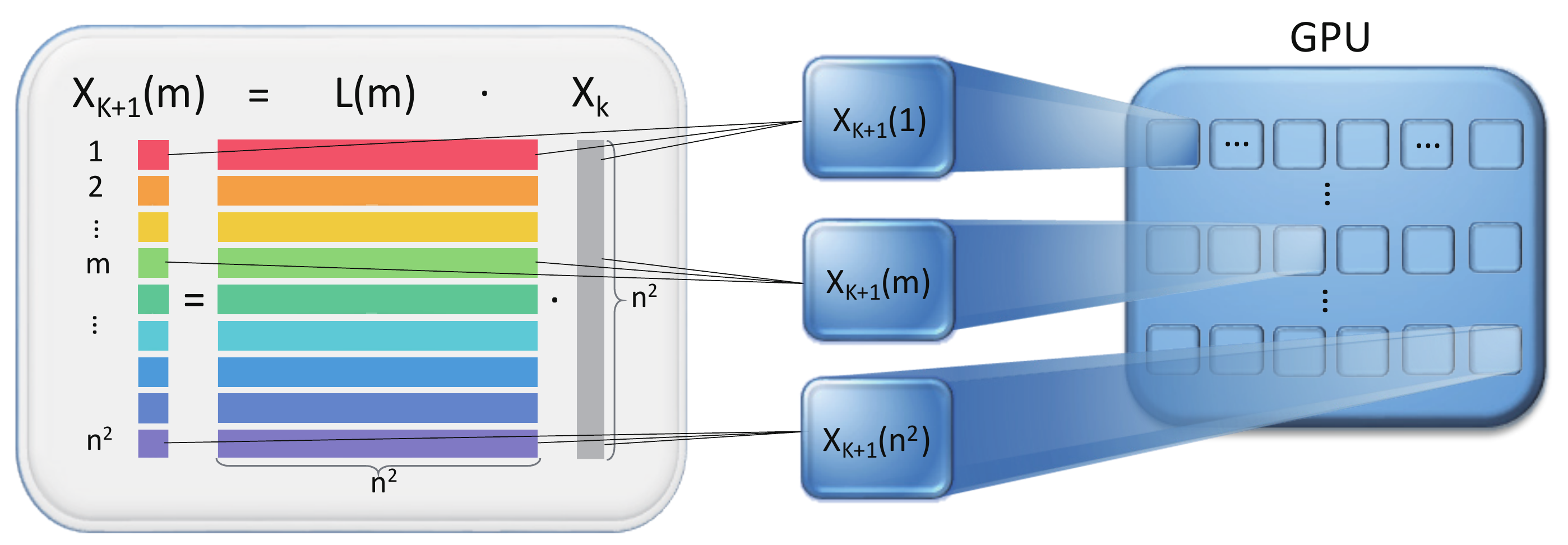}
\caption{\textbf{Schematic diagram of the solver framework.} The yellow box shows a normal example for a Kronecker product that suggests the expansion of matrix size. The grey framework shows the calculation mechanism of using Runge-Kutta method to convert a Kronecker product to a serial of separate calculations on each component of the matrix. The blue framework suggests the mechanism of GPU parallel computing.}
\label{fig:QFTConcept}
\end{figure*}

In recent years, quantum physicists introduce a quantum protocol of PageRank and hope to bring quantum advantages to the classical PageRank \cite{Paparo2012}. As the internet based on quantum computers is currently not available, quantum PageRank now mainly focuses on its implementation on a classical computer with consideration of quantum mechanics\cite{Paparo2012}. It essentially replaces the classical random walks in Google PageRank with quantum stochastic walks\cite {Sanchezburillo2012}. Such a model of flexibly mixing classical random walk and quantum walk \cite{Whitfield2010} already has wide applications in energy transport problems\cite{Caruso2016}, associative memory in Hopfield neural networks\cite{Schuld2014, Tang2019}, decision making\cite{Martinez2016}, and more issues in open quantum systems. Applying quantum stochastic walk to quantum PageRank, a few advantages over classical PageRank have been demonstrated\cite {Sanchezburillo2012}, for instance, to generate more accurate ranking by reducing degeneracy from elements of the same probability, and to have better notification of the significance of secondary-hubs in the network, etc.  

However, there lies a very severe challenge for numerical calculations of quantum stochastic walks, especially in the application of quantum PageRank that normally tackles large-scale networks. Quantum stochastic walk is based on a Lindblad master equation \cite{Lindblad1976} and involves lots of Kronecker products in its form. For a network with $N$ nodes, its Hamiltonian and Lindblad matrix are both in $N \times N$ dimension, and then the numerical calculation for Kronecker products would involve a dimension of $N^4$. This can cause huge explosion of required memory and exhaust a laptop by a network of 100-150 nodes, while a common network in real life normally exceeds that scale. There have currently been a number of solvers for quantum stochastic walk, including $Qutip$ \cite {Johansson2012, Johansson2013}, a Python package, and $QSWalk$ \cite{Falloon2017}, a Mathematica package that's specifically tailored for tasks on directed graphs. However, all current packages can only solve quantum stochastic walk of a limited scale, and none employs good enough optimization algorithms to level up the numerical capability for large-enough networks with above even hundreds of elements.

\begin{figure*}[ht!]
\includegraphics[width=1\textwidth]{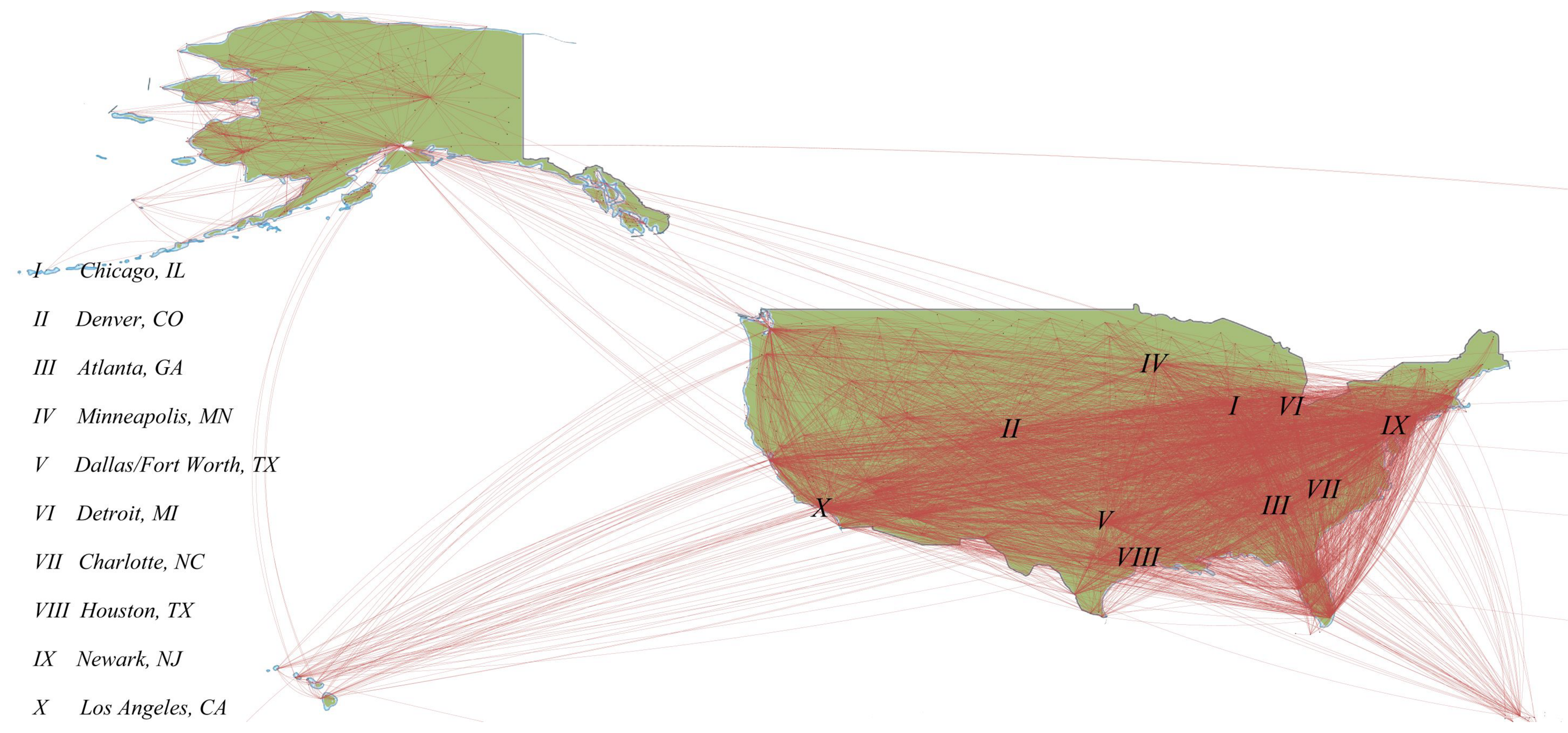}
\caption{\textbf{The USA airline network.} The airlines among major USA airports are plotted in the map using red lines. The top 10 important airports suggested by quantum PageRank are marked in the map.   
}
\label{fig:1Dexperiment}
\end{figure*}

Fortunately, there are ways to improve. Firstly, the Runge-Kutta method has long been used to solve ordinary differential equations\cite{Moler1978, Ehle1975, Shampine1976, Mathews2004}. It essentially reduces the matrix dimension to improve the calculation, which can also be applied to solve the quantum stochastic walks with a much smaller memory requirement. Besides the numerical method, the emerging TensorFlow framework and GPU parallel computing can also be utilized. TensorFlow is an open-source software library developed by Google for dataflow programming\cite{TensorFlow2020}. TensorFlow is useful for operating machine learning applications such as neural networks on multidimensional data arrays (tensors), and now it's also used to solve physics problems\cite{Cai2018, Carrasquilla2017, Broecker2017, Swaddle2017}. TensorFlow can run on not only CPU, but also on GPU, the graphics processor unit, and operates GPU parallel computing. GPU computing in the last decade has been developed rapidly. Through efforts by companies like NVIDIA, GPUs improve much faster than CPUs, reaching a large capacity of billions of transistors, and in the meantime, the work mechanism of GPUs to quickly create, run and retire multi-threads makes parallelism an inherent advantage for GPUs. GPU parallel computing has benefited many general purpose scientific computational problems by bringing up significantly speed-up performances\cite{Harris2005, Luebke2006, Leung2017}. 

Therefore, in this Letter, we present an efficient solver for large-scale quantum PageRank using the Runge-Kutta method to reduce the matrix dimension to O($N^2$) and employing TensorFlow to conduct GPU parallel computing. We demonstrate its performance in solving quantum PageRank for the USA major airline network with up to 922 nodes. Compared with the previous quantum PageRank solver, our solver dramatically reduces the required memory and time to only 1\% and 0.2\%, respectively, making it practical to work in a normal computer with a memory of 4-8 GB in just a few seconds. This efficient solver for large-scale quantum PageRank and quantum stochastic walks would greatly facilitate the study of quantum information in real-life applications. 

%\section{Main}
As has been mentioned, Google PageRank uses a very straightforward model with classical random walk and quantum PageRank essentially replaces it with quantum stochastic walk\cite {Sanchezburillo2012}. The detailed model for both PageRank protocols have been explained in Supplementary Note 1. The task for improving this solver can boil down to the problem of solving the quantum stochastic walk that is normally expressed in the Lindblad master equation: 

\begin{equation}
\begin{split}
 \frac{d\rho}{dt}=-(1-\omega)i[H,\rho(t)]+\\\omega\sum_{k=1}^{K}(L_k\rho(t)&L_k^{\dagger}-\frac{1}{2}(L_kL_k^{\dagger}\rho(t)+\rho(t)L_k^{\dagger}L_k)
\end{split}
\end{equation}
where $\rho$ is the density matrix that needs to be solved since it works out the element ranking. The parts with Hamiltonian $H$ and Lindblad terms $L$ describe the quantum walks and classical random walk, respectively. 

For this equation, we can rewrite it in such a form: $\frac{d\tilde{\rho}}{dt}=\mathscr{L}\cdot\tilde{\rho}(t)$, where $\tilde{\rho}$ is the transpose of the matrix $\rho$. Then $\rho$ can be solved by matrix exponential method: $\tilde{\rho}(t) = e^{\mathscr{L}t}\cdot \tilde{\rho}(0)$. The expression for $\mathscr{L}$ reads as follows:

\begin{equation}
\begin{split}
 \mathscr{L}=-(1-\omega)i(I_N\bigotimes H-H^T\bigotimes I_N)+\\ \omega\sum_{k=1}^K(L_k^{\dagger}\bigotimes L_k-\frac{1}{2}(I_N\bigotimes L_k^{\dagger}L_k+&L_k^TL_k^*\bigotimes I_N))
\end{split}
\end{equation}
where $\bigotimes$ is the Kronecker product. 

This suggests, for a network with $N$ elements, the Hamiltonian $H$ and Lindblad matrix $L$ are all always of a size $N \times N$, and then the Kronecker product would result in a size of $N^2 \times N^2$. If we calculate Eq.(2) directly, the memory we will need is approximately 2$N^4$, which suggests, a computer with a memory of 8GB can hardly afford the calculation for a network with more than 150 nodes. 

However, we notice that for a Kronecker product $C_{n^2\times n^2}$ in the form  $C_{n^2\times n^2}=A_{n\times n}\bigotimes B_{n\times n}$, each element in $C$ can be calculated separately as follows:

\begin{equation}
\begin{split}
 C_{i,j}=A_{1+(i-1)div \ n,1+(j-1)div \ n}\times \\ B_{1+(i-1)mod \ n,1+(j-1)mod \ n} 
\end{split}
\end{equation}

\begin{figure*}[hbt!]
\includegraphics[width=0.8\textwidth]{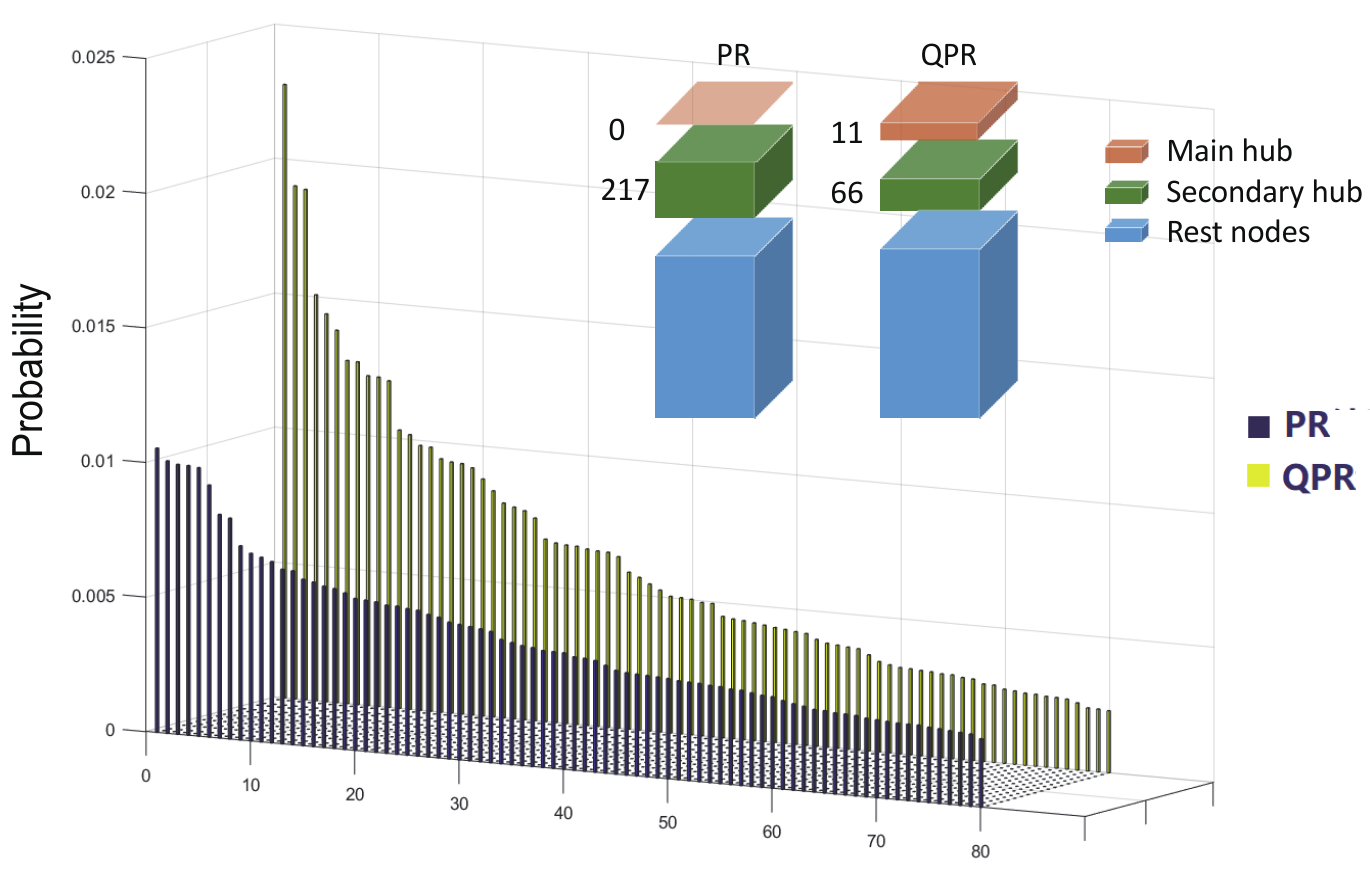}
\caption{\textbf{A comparison between classical PageRank and quantum PageRank.} The probability distribution for the the top 80 nodes using classical PageRank (PR) and quantum PageRank (QPR). Inset: the breakdown into categories of `main hub', `secondary hub' and `rest nodes' for PR and QPR. }
\label{fig:apparato}
\end{figure*}

Therefore, although $L$ is a matrix consisting $N^4$ elements, they each can be obtained from information of $O(N^2)$. When solving the equation 
    $\frac{d\tilde{\rho}}{dt}=\mathscr{L}\cdot\tilde{\rho}(t)$, we make use of the Runge-Kutta method, which does not require the storage of the whole matrix $L$. By using only one row of $L$ each time, we reduce the memory requirement down to approximately $100 N^2$. Details for Runge-Kutta method  are given in Supplementary Note 2.

In the meantime, as the calculation for each element of a matrix is independent, the calculation sequence for elements has no influence on the numerical results. Such a feature is exactly suitable for parallel computing, mapping the calculation for each element into one processing unit and calculating a large number of elements simultaneously. The GPU parallel computing is conducted using TensorFlow and we also run CPU computing in TensorFlow as a comparison. The framework of our solver of using Runge-Kutta numerical method and TensorFlow GPU parallel computing is illustrated in Fig. 1. 

Having constructed the solver, we would demonstrate it for quantum PageRank on a large-scale network in real-life: the highly developed USA airlines among 922 main USA airports, $i.e.$ a network of 922 nodes. The data of airline and airport information is retrieved from the USA Department of Transportation \cite{}. As shown in Fig. 2, we plot some airports and airlines in the USA map according to their real longitude and latitude information. For visual effects, we just plot 80 airports that have largest number of airlines and their airlines, instead of all 922 airports and over 14000 airlines that would otherwise make the figure a mess. Even Fig. 2 just shows partial network, we can still see the network covers all states across the USA and flights are regarded as a most important option for transportation in USA. Therefore, information of significance ranking for these airports would be of great meaning for the country. 

We use the solver to import the raw data that shows all airlines departing from an airport and arriving at another airport as the connection profile. Note that for many other networks, for instance, the hyperlinks for a company website, the connection profile cannot be directly downloaded in a spreadsheet. Instead, we can use the web scraper code to get the information. This is not included in quantum PageRank solver, but we also provide this web scraper code separately in Supplementary Data to facilitate users. Now the quantum PageRank solver can generate the Hamiltonian matrix and Lindblad matrix from the connection profile. The solver then employs the Runge-Kutta method with adaptive step length, termed as $RKF45$ method\cite{Mathews2004}, and loads TensorFlow GPU parallel computing. The outcome is the probability distribution at all elements. The ranking of these probability suggests the ranking of significance of these elements, in this case study, the ranking for all 922 USA airports. The whole code for this solver is given in Supplementary Data. 

\begin{figure*}[hbt!]
\includegraphics[width=1.0\textwidth]{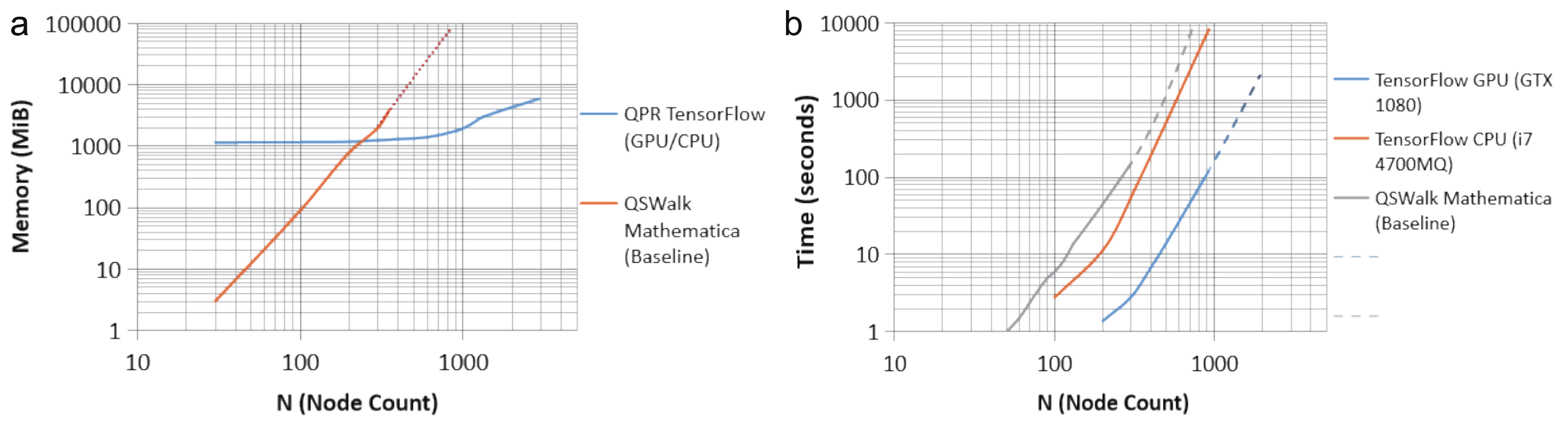}
\caption{\textbf{Calculation performance.} Probability distributions for ({\bf a}) Scaling of the required calculation time as a function of the node count $N$ using different solutions. The GPU and CPU used in the calculation are NVIDIA $GTX 1080$ and $i7 4700MQ$, respectively. The result for $QSWalk$ Mathematica solver is retrieved from reference \cite{Falloon2017}.  The short-dashed line is a prediction from the obtained scaling behavior.   ({\bf b})the required time.  Scaling of the required calculation memory as a function of the node count N using different solutions. The result for $QSWalk$ Mathematica solver is retrieved from reference \cite{Falloon2017}.  The short-dashed line is a prediction from the obtained scaling behavior.   
}
\label{fig:Results4}
\end{figure*}

We can now analyze the obtained quantum PageRank results. The solver gives the top-10 airports according to the ranking of their probabilities: Chicago, IL; Denver, CO; Atlanta, GA; Minneapolis, MN; Dallas/Fort Worth, TX; Detroit, MI; Charlotte, NC; Houston, TX; Newark, NJ; Los Angeles, CA. Marking these airports in Fig. 2, it shows that they are indeed the most popular USA airports that serve as the hubs to link to many airports through airlines.  

The result has also well spotted the `secondary-hubs', which are those in the tier-two ranks with still many connections with other airports in certain region. A quantitative definition of the `secondary-hub' has been given \cite{Paparo2012}: a node with its probability larger than $1/N$ and smaller than $c/N$, where $c$ is a constant set to be 10, and similarly, a `main-hub' is a node with a probability larger than $c/N$. In the case of USA airport ranking, for comparative studies, we have conducted a classical PageRank Gephi using $Gephi$, an open-source network analysis software\cite{Gephi2020}. Classical PageRank suggested that there are 217 nodes above $1/N$ and none above $c/N$. In other words, it does not spot the main hubs and regards up to 217 airports as the secondary hubs, which seems not a very realistic classification. For instance, some airports that  only connects to two other nodes are still regarded as the secondary hubs according to classical PageRank. On the other hand, our quantum PageRank gives better classification of the main hubs and secondary hubs: 11 nodes are above $c/N$ and 66 nodes between $1/N$ and $c/N$ (See inset of Fig. 3). A reason for the sharper identification of secondary hubs by quantum PageRank can be found from the obtained probability distribution by the two approached (Fig. 3). Classical PageRank has more averaged distribution over all nodes, and does not differentiate nodes of higher ranking with a probability separation as clear as the quantum PageRank. 

We further investigate the performance of this quantum PageRank solver in terms of the required time and memory. We demonstrate that for a network with a size of 1000 nodes, using the built-in $MatrixExp$ function in Mathematica\cite{Falloon2017}, it requires a memory of nearly 200GB , which is almost impossible on a single computer without the aid of high-performance computing. In fact, a size of 200-300 nodes is normally an upper bound for this solution on an averaged laptop. On the other hand, our approach only requires 2GB for $N$=1000, which can be easily implemented and greatly saves the computational resources. In our approach, the high-dimensional matrix goes under through an equivalent transformation to low-dimensional arrays, and hence the required memory drops to only 1\% of that for the previous Mathematica solution \cite{Falloon2017} (See Fig.4a). For the required calculation time, if running the built-in $MatrixExp$ function in Mathematica\cite{Falloon2017}, it would take 14 hours under the assumption that there is no limit in the memory . It takes around 3 hours for our TensorFlow CPU solution to finish the task on a common laptop (CPU configuration: $i7 4700MQ$), and around 1 hour on a desktop CPU with higher configurations.   On the other hand, it takes no more than 2 minutes using our TensorFlow GPU solution by taking full advantages of the parallel computing in GPU units. This is only 0.2\% of the time for finishing the same task using the Mathematica solution (See Fig.4b).

%\section{Discussion}
In this work, we have demonstrated an efficient quantum PageRank solver and showed its impressive performance in a case study of a very common network in real life. The solver requires only 1\% of the memory and 0.2\% of the time for this task comparing to working it on the previous Mathematica quantum PageRank solver, so that now users can easily conduct the ranking for large-scale networks in a normal computer for minutes instead of in a hyper-performance workstation for hours or days. This is of great meaning, as real-life networks such as Internet, transportation network and bacteria groups are always of high complexity and a large number of nodes. Only when we overcome the hurdle of calculating large-scale networks, can we really apply this useful and precise quantum PageRank method to practical ranking problems.  

As the core of this solver is to improve the calculation for Lindblad master equations, the important equation used to analyze quantum stochastic walks, we could as well apply this solver to many more problems described by quantum stochastic walks apart from quantum PageRank. Potential scenarios may include the simulation for open quantum systems in neural networks, photosynthetic process, and so on. They may also take advantages of the efficient calculation on large-scale samples.

Furthermore, the idea of using Runge-Kutta method to reduce the matrix size in order to reduce the memory is not new but still very useful for solving even more quantum physics problems in the future, since the matrix calculation is always a main task in quantum mechanics. Besides, the method of using TensorFlow GPU parallel computing to reduce the time is a powerful approach with strongly growing popularity. Though it is now more commonly discussed for machine learning, and only has few implementations for physics problems, it is highly suggested to combine this emerging powerful tool with advanced numerical methods to together contribute to the large variety of complicated quantum tasks. Such multidisciplinary research would boost the research for quantum information science greatly. 

\section*{Acknowledgements} 
The authors thank Jian-Wei Pan for helpful discussions. This research was supported by the National Key R\&D Program of China (2019YFA0308700, 2017YFA0303700), the National Natural Science Foundation of China (61734005, 11761141014, 11690033), the Science and Technology Commission of Shanghai Municipality (STCSM) (17JC1400403), and the Shanghai Municipal Education Commission (SMEC) (2017-01-07-00-02- E00049). H. T. is supported by National Natural Science Foundation of China (NSFC) (11904229), China Postdoctoral Science Foundation (19Z102060090). X.-M.J. acknowledges additional support from a Shanghai talent program.

\clearpage
\newpage

%%%%%%%%%%%%%%%%% Supplemental Information %%%%%%%%%%%%%%%%%%%%
\onecolumngrid
\section*{\large Supplemental Information: TensorFlow Solver for Quantum PageRank in Large-Scale Networks}
\setcounter{figure}{0}
\setcounter{table}{0}
\setcounter{equation}{0}
\renewcommand{\figurename}{Supplementary Figure}
\renewcommand{\tablename}{Supplementary Table}

\renewcommand{\thetable}{\arabic{table}}
\renewcommand{\theequation}{{S}\arabic{equation}}

\bigskip
\section*{\large Supplementary Note 1: The model for Google PageRank and Quantum PageRank}
\subsection{The model for Google PageRank}
Suppose we have a graph of $N$ nodes, then an $N \times N$-dimensional adjacency matrix $A$ can be used to describe the connections between the $N$ nodes. If there's a connection from Node $j$ to Node $i$, then $A_{ij}=1$, and otherwise $A_{ij}=0$. If all connections are undirected, $A_{ij}=A_{ji}$ applies for all $i$, $j$, and this is an undirected graph. If a network has directed connections, then $A_{ij}\not=A_{ji}$. We further define the out-degree as sum of connections leaving a Node $j$: $outDeg(j)=\sum A_{ij}$. 

As we have mentioned, the Google PageRank algorithm is essentially a small adaption from the classical random walk, which we'll introduce here briefly. For a continuous-time classical random walk[Ref S1], the probability evolution has a relationship with the transition matrix $M$ and out-degree: $\frac{d\rho}{dt}=M\cdot\rho(t)$, where
$\frac{d\rho}{dt}=G\cdot\rho(t)$, where
\begin{equation}
M_{ij}=
\begin{cases}
-A_{ij},~~{\rm if} ~i\not=j\\
-outDeg(j),~~{\rm if} ~i=j
\end{cases}
\end{equation}

\begin{equation}
F_{ij}=
\begin{cases}
1/(N-1),~~{\rm if} ~i\not=j\\
0,~~{\rm if} ~i=j
\end{cases}
\end{equation}

Then the probability distribution $\rho$ can be obtained by:
\begin{equation}
\rho(t) = e^{-Mt}\cdot \rho(0)
\end{equation}

For Google PageRank, a Google matrix $G$ [Ref S2] is used to adapt from the transition matrix $M$: 
\begin{equation}
G=qM+(1-q)F
\end{equation}

where $q$ is typically set as 0.9, and $F$ is the long-distance hopping matrix, $F_{ij}=1/(N-1)$ if $i\not=j$; $F_{ij}=0$ if $i=j$. By doing so, there is no zero terms in the matrix $G$ and the long-term probability would not be an even distribution in all nodes as in the case of using matrix $M$. We put the matrix $G$ in Eq.(S1) and we can get the probability distribution as the criteria for element ranking. 

\subsection{The model for Quantum PageRank}
Now let's come to the model of Quantum PageRank. Same to have a long distance hopping matrx $F$, the Quantum PageRank matrix $Q$ reads as:

\begin{equation}
Q=q \mathscr{L}+(1-q)F
\end{equation}

where $\mathscr{L}$ represents the transition matrix for quantum stochastic walks, which is a combination of classical random walk and quantum walk. The expression of  $\mathscr{L}$ is given in Eq.(2) in the main text.

A continuous-time quantum walk has many similarities with the classical random walk. The probability is replaced by the quantum state vector $\ket{\Psi(z)}$  whose mode is the probability for each node, and the Hamiltonian matrix for quantum walks $H$ has the similar definition with $M$. Therefore, the state vector evolution can be similarly defined by a Schr\"odinger equation with a solution: $\ket{\Psi(z)}=e^{-iHz}\ket{\Psi(0)}$. 

The pure quantum walk still has a clear difference from classical random walks in terms of the symmetry of matrix $M$ or $H$. Due to the requirement of unitary evolution of quantum walks, $H$ must be Hermitian and hence is symmetric.

\begin{equation}
H_{ij}=
\begin{cases}
-max(A_{ij}, A_{ji}),~~{\rm if} ~i\not=j\\
-outDeg(j),~~{\rm if} ~i=j
\end{cases}
\end{equation}
This means $H$ always satisfies: $H_{ij}=H_{ji}$, and hence pure quantum walk does not apply to directed graphs where some  $A_{ij}\not=A_{ji}$, while classical random walk does not have such restrictions as $M_{ij}$ can be asymmetric.
 
Fortunately, the quantum stochastic walk can be well utilized for both undirected and directed graph, as well as taking into consideration of the quantum evolution. It uses the Lindblad equation as it main form [Ref S1]:  
\begin{equation}
\frac{d\rho}{dt}=-(1-\omega)i[H,\rho(t)]+\omega\sum_{k=1}^{K}(L_k\rho(t)L_k^{\dagger}-\frac{1}{2}(L_kL_k^{\dagger}\rho(t)+\rho(t)L_k^{\dagger}L_k)
\end{equation}
where the probability distribution $\rho$ is related to a mixture of the quantum walk (the part containing $H$) and the classical random walk (the part containing $L$ and $L^{\dagger}$), and  $\omega$ interpolates the weight for the two kinds of walks. $H$ is always symmetric with $H_{ij}=H_{ji}$ in order to satisfy the Hemitian. On the other hand, the directed connection can be reflected by the Lindblad terms, e.g. $L_{ij}$ corresponds to a specific scattering channel from Node $j$ to Node $i$: $L_{ij}=\sqrt{|M_{ij}|}\ket{i}\bra{j}$. If there's a connection from Node $j$ to Node $i$, $L_{ij}\not=0$, and vise versa, $L_{ij}=0$. Therefore, all those non-zero $L_{ij}$s present the full profile including both undirected and directed connections within the graph.  

Rewriting the Lindblad equation in Eq. (S6) in such a form: $\frac{d\tilde{\rho}}{dt}=\mathscr{L}\cdot\tilde{\rho}(t)$, we can get the transition matrix $\mathscr{L}$ for quantum stochastic walks, and we can then apply it to the model of quantum PageRank. 

\newpage

\section*{\large Supplementary Note 2: Explanation for the Runge-Kutta Method}

As has been mentioned in the main text, the matrix that we need to conduct matrix exponential method is as follows: 
\begin{equation}
   \mathscr{L}=-(1-\omega)i(I_N\bigotimes H-H^T\bigotimes I_N)+\omega\sum_{k=1}^K(L_k^{\dagger}\bigotimes L_k-\frac{1}{2}(I_N\bigotimes L_k^{\dagger}L_k+L_k^TL_k^*\bigotimes I_N))
\end{equation}

It is a matrix with the size of $N^2 \times N^2$, so if we calculate the matrix exponential in the classic way, the memory we will need is approximately 2$N^4$, $i.e.$, a computer with a memory of 8GB can hardly afford the calculation for a network with above 150 nodes. However, we notice that $\mathscr{L}$ is the sum of several kronecker products. For  $C_{n^2\times n^2}=A_{n\times n}\bigotimes B_{n\times n}$, each element in $C$ can be calculated as follows:

\begin{equation}
 C_{i,j}=A_{1+(i-1)div \ n,1+(j-1)div \ n}\times B_{1+(i-1)mod \ n,1+(j-1)mod \ n}  
\end{equation}

Therefore, although $L$ is a matrix consisting $N^4$ elements, they each can be obtained from information of $O(N^2)$. This is a very suitable task to take advantage of the Runge-Kutta numerical method. 

We use the Runge-Kutta method with an adaptive step size $h$, denoted the RKF45 method[Ref S3]. It works as follows: let $y_0$ be the initial state, and we can obtain the final state $y_t$ ($t=t_i/h$, where $t_i$ is the evolution time of quantum random walk, and $h$ is the step size of the iteration.) by iterating using the following coefficients:

%$$ x_{j+1} = x_{j}+\frac{1}{6}k_1+\frac{1}{3}k_2+\frac{1}{3}k_3+\frac{1}{6}k_4$$

%$$ k_1=hLx_j,k_2=hL(x_j+\frac{1}{2}k_1),k_3=hL(x_j+\frac{1}{2}k_2),k_4=hL(x_j+k_3)$$

%\begin{spacing}{2.0}
%\linespread{2.0}
\begin{equation}
\begin{aligned}
&k_1 = hf(t_k, y_k)\\
&k_2 = hf(t_k+\frac{1}{4}h, y_k+\frac{1}{4}k_1)\\
&k_3 = hf(t_k+\frac 3 8 h, y_k+\frac{3}{32}k_1+\frac{9}{32}k_2)\\
&k_4 = hf(t_k+\frac{12}{13}h, y_k+\frac{1932}{2197}k_1-\frac{7200}{2197}k_2+\frac{7296}{2197}k_3)\\
&k_5 = hf(t_k+h, y_k+\frac{439}{216}k_1-8k_2+\frac{3680}{513}k_3-\frac{845}{4101}k_4)\\
&k_6 = hf(t_k+\frac{1}{2}h, y_k-\frac{8}{27}k_1+2k_2-\frac{3544}{2565}k_3+\frac{1859}{4104}k_4-\frac{11}{40}k_5)\\
\end{aligned}
\end{equation}
%\end{spacing}

Then an approximation of the value $y$ can be obtained using an iteration relationship with an order of 4:
\begin{equation}
y_{k+1} = y_k+\frac{25}{216}k_1+\frac{1408}{2565}k_3+\frac{2197}{4101}k_4-\frac{1}{5}k_5
\end{equation}

It can alternatively use a relationship with an order of 5. We use $z_{k+1}$ for this iteration relationship in order to differ from $y_{k+1}$ used in the above equation. 

\begin{equation}
z_{k+1} = y_k+\frac{16}{135}k_1+\frac{6656}{12,825}k_3+\frac{28,561}{56,430}k_4-\frac{9}{50}k_5+\frac{2}{55}k_6
\end{equation}

Note that there would be some discrepancy between the value for $y_{k+1}$ and $z_{k+1}$ that use 4-order and 5-order equation, respectively. A term $s$ is considered that's calculated as follows:

\begin{equation}
s = \left(\frac{\text{Tol}\ h}{2\left|z_{k+1}-y_{k+1}\right|}\right)^{1/4}\approx0.84\left(\frac{\text{Tol}\ h}{\left|z_{k+1}-y_{k+1}\right|}\right)^{1/4}
\end{equation}

Therefore, for an acceptable tolerance of discrepancy that we have set, denoted as Tol, we can update the step size by multiplying  $s$ with the current step size $h$, making $sh$ the updated step size. 

\section*{Supplementary References}

%\begin{adjustwidth}{1cm}{0cm}

\noindent [S1] Falloon, P., Rodriguez, J. \& Wang, J. \emph{QSWalk}: a Mathematica package for quantum stochastic walks on arbitrary graphs. \emph{Comput. Phys. Commun.} {\bf 217}, 162-170 (2017).

\noindent [S2] S\'anchez-Burillo, E., Duch, J., G\'omez-Garde\~nes, J. \& Zueco, D. Quantum navigation and ranking in complex networks. \emph{Sci. Rep.}
{\bf 2}, 605 (2012).

\noindent [S3] Mathews, J. H., \& Fink, K. K. \emph{Sec.9.5 Runge-Kutta Methods, Numerical Methods using Matlab, Fourth Edition.} Prentice-Hall Inc. 497-499 (2004).

%\begin{thebibliography}{10}
%\renewcommand{\bibnumfmt}[1]{#1.}
%\end{thebibliography}

\bigskip

\end{document}